\documentclass[%
reprint,
nofootinbib,
 amsmath,amssymb,
aps,
prl,
]{revtex4-1}

\usepackage{graphicx}
\usepackage{dcolumn}
\usepackage{bm}

\usepackage{tikz}
\usetikzlibrary{matrix,arrows,decorations.pathmorphing,decorations.pathreplacing}
\usepackage{euscript,array,mathrsfs}

\def\BA{A}

\def\BH{{\mathscr H}}

\newcommand{\Tr}{\operatorname{Tr}}
\newcommand{\IM}{\operatorname{Im}}

\newcommand{\GAP}{\phantom{~~~~~~~~~~~~~~~~~~~~~~~~~}}

\def\bra#1{\langle #1|}
\def\ket#1{| #1\rangle}

\newcommand{\EQ}[1]{\begin{equation}\begin{split} #1
\end{split}\end{equation}}

\newcounter{Part}
\newcommand{\Part}{\refstepcounter{Part}\vspace{0.5cm}{\bf \arabic{Part}. }}

\begin{document}

\preprint{APS/123-QED}

\title{New Modal Quantum Mechanics}

\author{Timothy J. Hollowood}
\email{t.hollowood@swansea.ac.uk}
\affiliation{
Department of Physics, Swansea University,\\ Swansea, SA2 8PP, United Kingdom
}

\date{\today}

\begin{abstract}
We describe an interpretation of quantum mechanics based on reduced density matrices of sub-systems from which the standard Copenhagen interpretation emerges as an effective description for macro-systems. 
The interpretation is a modal one, but does not suffer from the range of problems that plague other modal interpretations. 
The key feature is that quantum states carry an additional property assignment in the form of one the eigenvectors of the reduced density matrix which evolves evolves according to a stochastic process driven by the unmodified Schr\"odinger equation, but it is usually hidden from the emergent classical description due to the ergodic nature of its dynamics. However, during a quantum measurement, ergodicity is broken by decoherence and definite outcomes occur with probabilities that agree with the Born rule.
\end{abstract}

\maketitle



\Part It is a widely-held view that the interpretation of quantum mechanics is not worthy of study because the theory can be used to predict phenomena with incredible accuracy. According to this view, if the issue of interpretation needs any work then it must only be of a  modest amount that aims to put the Copenhagen interpretation (standard textbook quantum mechanics) on a firmer footing. So the collapse of the wave-function should not be put in by hand; rather, it must emerge in suitable situations from more fundamental rules. This letter describes an approach to these issues that makes a minimal change to rules of standard quantum mechanics: brains, minds, observers or many worlds, for that matter, are not needed. A more complete version appears in \cite{long}.

Standard quantum mechanics describes the evolution of the state vector via the Schr\"odinger equation. Whilst, we are content to describe a microscopic system by a state vector, state vectors that involve a superposition of macroscopically distinct states seem at odds with the classical world that must emerge.
In addition, when quantum phenomena manifest themselves in the emergent classical world, they do so in an apparently stochastic way even though the underlying state vector evolves smoothly.  

The central idea of the new interpretation is to overlay the standard description of the quantum state 
that evolves according to an unmodified Schr\"odinger equation 
with additional stochastic information but in a way that is consistent with the quantum violation of Bell's inequality.

\Part The quantum systems that we are usually concerned with involve localised systems that are interacting with a much larger environment just as in classical statistical mechanics where systems are considered to be interacting weakly with a large thermal bath. In the quantum case, the standard description of the quantum system $A$ is then via the reduced density matrix $\hat\rho_A$ obtained by taking the state of the $A$ plus the environment $E$, a pure state $\Psi$, say, and tracing out the degrees-of-freedom of $E$. It is important that, as long as the environment is very large, $d_E\gg d_A$, with $d_A=\text{dim}\,\BH_A$, etc., the actual state $\Psi$ is largely irrelevant. Evolution of $\hat\rho_A$ then follows from the Schr\"odinger equation for $A+E$. Recent  quantum approaches to classical statistical mechanics \cite{PopescuShortWinter:2006efsm,GoldsteinLebowitzTumulkaZanghi:2006ct,GMM} view $\hat\rho_A$ as the fundamental definition of an ensemble for which the thermodynamical entropy is identified with its entanglement entropy $-\Tr(\hat\rho_A\log\hat\rho_A)$. 

\Part In what follows a important role will be plays by a typical magnitude $\Delta$ of an inner product between two macroscopically distinct states. Just to get a feel for such a scale, let us estimate it by
supposing that macro-system has a macroscopic number of degrees-of-freedom $N\sim 10^{20}$. States are macroscopically distinct if all the microscopic degrees-of-freedom 
are separated by a macroscopic scale $L\sim 10^{-4}\,\text{m}$. If $\ell\sim 10^{-10}\,\text{m}$ is a characteristic microscopic length scale in the system and assuming, say, Gaussian wave-functions for the microscopical degrees-of-freedom spread over the scale $\ell$, the matrix elements between macroscopically distinct ontic states are roughly $\Delta\sim e^{-NL^2/\ell^2}\sim e^{-10^{32}}$. 

\Part The new interpretation lies firmly in the class of modal interpretations (see \cite{Krips:1969tpqm,vanFraassen:1972faps,Cartwright:1974vfmmqm,Bub:1992qmwpp,VermaasDieks:1995miqmgdo,BacciagaluppiDickson:1999dmi,Vermaas:1999puqm} and references therein) but does not suffer from the range of problems that plague existing versions \cite{Hollowood:2013cbr,long}. Like certain modal interpretations, the new one
pushes the analogy with classical statistical mechanics a bit further in that it postulates that a sub-system $A$ has in addition to its {\it epistemic state\/} $\hat\rho_A$ an {\it ontic state\/} $\psi_i$ 
that is one of the eigenvectors of $\hat\rho_A$:
\EQ{
\hat\rho_A\ket{\psi_i}=p_i\ket{\psi_i}\ .
\label{pdx}
}
Note that this is a discrete and finite set of states and $\sum_ip_i=1$. Note that the dynamics of the ontic states will not be defined continuously in $t$ and so we do not need to consider exact degeneracies.

The ontic state is to be viewed as the actual state that the system occupies in analogy with the micro-state of classical statistical mechanics. It is important that this property assignment cannot be taken as a global assignment: when we trace over states in the complement to $A$ we are restricting the view to $A$ and there is no guarantee that this perspective will be compatible with the perspective of another sub-system $B$ \cite{BD1,BH,Dk1}. 
In this regard we will have to argue how a recognisable  
classical world can emerge from these different perspectives. However, it is important that $\psi_i$ an ontic state of $A$ is precisely correlated with $\tilde\psi_i$ an ontic state of $E$, that we call the mirror, via the Schmidt decomposition:
\EQ{
\ket{\Psi}=\sum_i\sqrt{p_i}\ket{\psi_i}\ket{\tilde\psi_i}\ .
}

The eigenvalue $p_i$ can be interpreted in certain situations, as will emerge, as the probability that the system is actually in the $i^\text{th}$ ontic state at time $t$ but the more fundamental probability is the condition probability $p_{i|j}(t',t)$ that if the system was in the $j^\text{th}$ ontic state at time $t$ then at a later time $t'$ it is in the $i^\text{th}$ ontic state. It is a hypothesis that these probabilities are related to the eigenvalues $p_i$ via
\EQ{
p_i(t')=\sum_jp_{i|j}(t',t)p_j(t)\ .
\label{pc6}
}

The ontic states follow a stochastic process that is defined by the conditional probabilities. We make the hypothesis that it is completely defined at some ultra-violet scale $\eta\ll\tau$, where $\tau$ is a typical decoherence time scale, in terms of $p_{i|j}(t)\equiv p_{i|j}(t+\eta,t)$. Note that we must not take $\eta\to0$ because every quantum system has an intrinsic cut off scale, here the time interval $\eta$, beyond which the description breaks down and ceases to be unitary. It is unrealistic to specify the process at scales that are smaller than $\eta$.
The complete stochastic process can be built up out the $p_{i|j}$ if the process is Markov. In that case over a series of time steps $t_n=t+(n-1)\eta$ 
\EQ{
p_{j_N|j_1}(t_N,t_1)=\sum_{j_2,\ldots,j_{N-1}}\left[\prod_{n=1}^{N-1}p_{j_{n+1}|j_n}(t_n)\right]\ .
\label{stp2}
}
Consequently the stochastic process is a discrete-time Markov chain. 

Note that over physically relevant time scales $\tau$ the discretisation errors are of order $\eta/\tau $ and will, therefore, be negligible. However, the fact that the process is coarse grained at the scale $\eta$
avoids the problem of the continuous-time limit
involving rapid transitions, on a scale $\tau\Delta\ll\eta$, between macroscopically distinct states when their associated probabilities $p_i$ and $p_j$ try to cross but, more generically, ``repel" each other
 \cite{BDV,Donald1,Vermaas:1999puqm}. These {\it macro-flips\/} are clearly completely unphysical and they result from pushing an effective theory beyond its range of validity.
 
Even with the constraint \eqref{pc6}, there is some freedom in defining the stochastic process; however, our hypothesised process has a number of desirable features. The first point is that the dynamics is smooth on the scale $\eta$ meaning that ontic states can be labelled so that $\bra{\psi_i(t+\eta)}\psi_i(t)\rangle=\delta_{ij}+{\cal O}(\eta/\tau)$. The basic transitions are \cite{long,Hollowood:2013cbr}, for $i\neq j$,
\EQ{
p_{i|j}&=\frac{2\eta}\hbar\sqrt{\frac{p_i}{p_j}}\text{max}\,\Big[
\IM\,\bra{\psi_i}\bra{\tilde\psi_i}\hat H_\text{int}\ket{\psi_j}\ket{\tilde\psi_j}
,0\Big]\ ,
\label{sp2}
}
where $\hat H_\text{int}$ is the interaction part of the Hamiltonian that couples $A$ and $E$. This expression is valid to order $\eta/\tau$ and an exact expression is given in \cite{Hollowood:2013cbr,long} which can be shown to satisfy \eqref{pc6}.
Note that consistency requires that $\sum_{j\neq i}p_{j|i}\sim \eta/\tau\ll1$, for each $i$, which can be taken as a definition of the decoherence scale $\tau$. If this is not satisfied then that is a signal the effective quantum description has broken down and a more fundamental description should be used. 

As well as being a discrete-time Markov chain, the stochastic process has two important features. It only depends on the local coupling $\hat H_\text{int}$ between the corresponding pair of ontic states and their mirrors and it leads generally to a ergodic process where any state can reach any other state in a finite number of steps. However, there is also the possibility for the breaking of ergodicity when the states $\psi_i$ and $\psi_j$ become macroscopically distinct in which case $p_{i|j}$ is suppressed by a factor of order $\Delta$ due to the locality of $\hat H_\text{int}$. This is exactly what happens during a quantum measurement as we will see later.

\Part Macro-systems are characterised by the fact that, in general, they are in equilibrium with their environments. It has been shown that a sub-system $A$ interacting with a much larger environment will equilibrate, that is $\hat\rho_A$ will end up fluctuating around a constant (more precisely a slowly varying equilibrium) whatever the initial state \cite{2009PhRvE..79f1103L}. In particular, this means that the underlying stochastic process becomes, approximately, a homogeneous discrete-time Markov chain.
This equilibrium process is generally  {\it ergodic\/} since any ontic state can reach any other ontic state in a finite number of steps. This means that after the characteristic time scale $\tau$ the $p_{i|j}(t+\tau,t)$ become independent of $j$ and so the slowly varying $p_i$ are interpreted as the probability of $A$ being in the $i^\text{th}$ ontic state regardless of the initial state:
\EQ{
\text{equilibrium:}\qquad p_{i|j}(t+\tau,t)\approx p_i(t+\tau)\ .
}
As time evolves the ontic states evolve ergodically through the equilibrium ensemble so that the time-averaged behaviour on macroscopic scales, what is effectively the emergent classical state, is captured by the ensemble average with respect to $\hat\rho_A$. 
So, for a macro-system in equilibrium, the $p_i$ can be given an ignorance interpretation from the point-of-view of the emergent classical description.
The importance of this is that the details of the ultra-violet stochastic process are largely irrelevant for a macro-system.
However, the measurement process described later shows that this is not always true because in these situations
ergodicity is broken and then the ontic state of the measuring device determines the outcome.

\Part The new interpretation cannot generally specify joint probabilities to ontic states of 2 disjoint sub-systems $A$ and $B$ since there is no guarantee that the 2 perspectives are compatible. The only information that we can have on both $A$ and $B$ are the ontic states $\Phi_m$ of the composite system $A+B$ and these need bare no relation to those of $A$ and $B$ separately.
However, if $A$ and $B$ are 2 macro-systems weakly-interacting or causally separated, embedded in a larger environment we expect that a classical picture can emerge in the sense that the ontic states of $A+B$ are to a high accuracy
of the form of tensor products of ontic states of $A$ and $B$:
\EQ{
\ket{\Phi_{m(i,a)}}=\ket{\psi_i}\ket{\phi_a}+{\cal O}(\Delta)\ ,
}
where $m(i,a)$ is a 1-to-1 map. In this case, a joint probability for ontic states of $A$ and $B$ can be said to have emerged: 
\EQ{
p(i,a)\overset{\text{emergent}}=p_{m(i,a)}\ .
\label{e44}
} 
Note that $A$ and $B$ will be classically correlated if $p(i,a)\neq p_ip_a$.

So a classical ontology can emerge from a set of macro-systems embedded in a larger environment by a patching together of ontic states of composite systems at order $\Delta$. Note, however, that there will be a spectrum of classicality set by the magnitude of $\Delta$. 

\Part Consider a measurement of $\sigma_z$ of a qubit. For a suitable Hamiltonian the solution of the Schr\"odinger equation gives
\EQ{
\big(c_+\ket{z^+}+c_-\ket{z^-}\big)\ket{\Phi_0}\rightarrow c_+\ket{z^+}\ket{\Phi_+}+c_-\ket{z^-}\ket{\Phi_-}\ .
}
Here, the states $\Phi_0$ and $\Phi_\pm$ are states of the measuring device $A$ plus environment $E$ before and after the measurement at $t=0$ and $T$. There are 3 relevant reduced density matrices of $A$
\EQ{
\hat\rho_A^{(i)}=\Tr_E\ket{\Phi_i}\bra{\Phi_i}\ ,\qquad i\in\{0,\pm\}\ .
}
Here, $\hat\rho_A(0)=\hat\rho_A^{(0)}$ is the initial density matrix, while 
\EQ{
\hat\rho_A(T)=|c_+|^2\hat\rho_A^{(+)}+|c_-|^2\hat\rho_A^{(-)}\ .
}
The ontic states of $A$ in the final state split up into two subsets ${\cal E}^{(\pm)}$ (as well as a set of approximately null eigenvectors $p_a\sim\Delta$ which play no important role) with 
\EQ{
\hat\rho^{(\pm)}_A=\sum_{a\in{\cal E}^{(\pm)}}p_a^{(\pm)}\ket{\psi_a(T)}\bra{\psi_a(T)}+{\cal O}(\Delta)\ .
}
It is important that $p_{a|b}$ in \eqref{sp2} for $a\in{\cal E}^{(\pm)}$ and $b\in{\cal E}^{(\mp)}$ are suppressed by a factor order $\Delta$ and
 so ergodicity is effectively broken in the final state. Hence, the measurement has a definite outcome depending on which sub-ensemble the final ontic state lies in.

In order to compute the probabilities of each of the outcomes $\pm$ we must use the stochastic process
to compute the probability of going from an initial ontic state to any final ontic state. Since the ontic state is hidden from the emergent classical description and assuming that $A$ is in equilibrium with $E$, we should average the initial state over the ensemble $\hat\rho_A(0)$ and then sum over either of the sub-ensembles ${\cal E}^{(\pm)}$ in the final state. The key point is that because of this averaging and \eqref{stp2}, the sum over histories is trivial and the probability of the outcome $\psi_a(T)$ is then just equal to:
\EQ{
\sum_bp_{a|b}(T,0)p_b(0)=p_a(T)\ .
}
So, up to corrections of order $\Delta$, the probability to be in the one of of the two sub-ensembles is
\EQ{
p^{(j)}(T)=\sum_{a\in{\cal E}^{(j)}}p_a(T)=|c_j|^2+{\cal O}(\Delta)\ ,
\label{bfr}
}
where we used the fact that $p_a(T)=|c_j|^2p_a^{(j)}$, for $a\in{\cal E}^{(j)}$, and $\sum_ap_a^{(j)}=1$, to order $\Delta$. So we have derived the Born rule.

After the measurement, once a particular outcome has emerged, say $\pm$, the breakdown of ergodicity means that it is prudent, but ultimately unnecessary, to remove the ergodically inaccessible component $\hat\rho_A^{(\mp)}$; this innocuous process piece of book keeping is effectively the collapse of the state vector
\EQ{
\hat\rho_\BA(T)~\rightsquigarrow~
\hat\rho_\BA^{(+)}~\text{ or }~\hat\rho_\BA^{(-)}\ .
}
In addition, the final ontic state of $A+\text{qubit}$ are
\EQ{
\ket{z^\pm}\ket{\psi_a(T)}\ ,\qquad a\in{\cal E}^\pm\ ,
}
which shows how the projection postulate emerges.

The whole discussion here can be generalised to include the more realistic case of a measuring device that is not perfect and can make errors and also measurements on continuous quantum systems \cite{Hollowood:2013cbr,long}. It can also be generalized to the case of measuring a continuous quantum systems like the position of a particle. Existing modal interpretations 
have had problems in situations like this with macroscopic superpositions of states \cite{B2,Page:2011gi}. However,
once one builds in realistic finite resolution scales no macroscopic superpositions occur \cite{Hollowood:2013cbr}.

\Part In this letter we have introduced a new and completely self-contained modal interpretation of quantum mechanics that we call the emergent Copenhagen interpretation described in more detail in \cite{long}. 

The key feature of the new representation is the definition of an underlying stochastic process that is defined at the ultra-violet scale $\eta$, although its actual definition is largely irrelevant as long as it satisfies some basic conditions. The existence of this allows for a solution of the measurement problem via an ergodicity argument in much the same way that symmetry is broken by a phase transition in classical statistical mechanics. In addition, a classical limit emerges by patching together ontic states of macro-systems embedded in a larger environment. The new interpretation also implies the standard quantum mechanical violation of Bell's inequality but in a way that preserves the locality of sub-systems \cite{long}.

\GAP

I would like to thank Jacob Barandes for a very fruitful exchange of ideas; many of my ideas were formed and are much sharper as a result of our communications. I am supported in part by the STFC grant ST/G000506/1.

\end{document}